# Andlantis: Large-scale Android Dynamic Analysis

Michael Bierma[†‡], Eric Gustafson[‡], Jeremy Erickson[†], David Fritz[†], Yung Ryn Choe[†]
[*†]Sandia National Laboratories
{mbierma, jericks, djfritz, yrchoe}@sandia.gov
[‡]University of California, Davis
{mhbierma, edgustaf}@ucdavis.edu

*Abstract*—

**Analyzing Android applications for malicious behavior is an important area of research, and is made difficult, in part, by the increasingly large number of applications available for the platform. While techniques exist to perform static analysis on a large number of applications, dynamic analysis techniques are relatively limited in scale due to the computational resources required to emulate the full Android system to achieve accurate execution. We present Andlantis, a scalable dynamic analysis system capable of processing over 3000 Android applications per hour. During this processing, the system is able to collect valuable forensic data, which helps reverse-engineers and malware researchers identify and understand anomalous application behavior. We discuss the results of running 1261 malware samples through the system, and provide examples of malware analysis performed with the resulting data.**

## I. INTRODUCTION

As the marketshare and popularity of the Android platform has expanded, the number of applications available to users has increased dramatically. It is estimated that 70% of the worlds smartphones run Android, with over 900,000 applications available on the official Google application store as of May 2013 [1]. This has made Android a tempting target for malware authors, and is the victim of 92% of mobile malware threats [2].

In an attempt to get a better understanding of the nature of this large volume of executable code, researchers employ a combination of static and dynamic analysis techniques. Static analysis—the analysis of a program's source or byte code to determine behavior—is much easier to scale, due to its reliance on data operations that are easily accelerated and parallelized. Previous static methods have scaled to hundreds of thousands of Android applications[12]. However, dynamic analysis—observing the program's execution to determine behavior—is much harder to scale due to the need to accurately replicate the desired execution environment, usually through the use of emulation or virtualization. In many cases, dynamic analysis is necessary to uncover vulnerabilities too complicated for static analysis, or discover flaws in program logic only known at runtime (*e.g.* dynamic dispatch). In addition to simply emulating the applications, the researcher must also obtain enough data from the running applications to detect the desired behaviors, adding an extra level of complexity to an already large computational burden.

In this paper, we present Andlantis: a highly scalable dynamic analysis framework for analyzing applications on the Android operating system. Andlantis runs the Android operating system in a virtualized environment and is able to provide the virtual device with artificial network data in order to provide an environment which closely replicates that of a physical device. Andlantis is able to schedule and run thousands of Android instances in parallel, enabling us to investigate the behavior of mobile malware at scale.

Andlantis employs a scalable high-performance emulytics framework, minimega, to parallelize this expensive task as much as possible and achieve a level of throughput unprecedented in Android dynamic analysis. Minimega provides the ability to coordinate and distribute our analysis across a cluster constructed from inexpensive, commodity hardware. Our experiment framework captures an application's disk and network activity, which is further parsed for anomalous behavior. Additionally, we are able to run applications within different operating environments and compare the results for interesting behavioral differences.

The rest of this paper provides: a description and analysis of related work, a detailed description of the Andlantis architecture, the results of our experiments on Android malware using Andlantis, an evaluation of the Andlantis system, a description of the limitations and future work, and a concluding section.

## II. RELATED WORK

To date, there have been a number of frameworks designed for the analysis of Android applications. We can classify these as static analysis [13], [9], [10], [18] and dynamic analysis [8], [5], [7], [19], [17], [15], [6], [4], [14] frameworks.

In [18], Schmidt et. al. utilize machine learning techniques to identify malicious binaries in Android applications through the extraction of functions and function calls found in the ELF binaries.

RiskRanker and Andarwin[13], [9] are scalable frameworks which provide valuable insight to Android application behavior. RiskRanker utilizes a variety of static analysis techniques to detect malware on Android devices. These techniques include program control flow graph evaluation and evaluation of byte code signatures. AnDarwin is designed for the scalable detection of plagiarized applications. This is performed through the construction of a Program Dependency Graph (PDG) of the Android application. A semantic vector is then extracted from the PDG to represent the application, which can then be quickly compared with the semantic vectors of other applications.

These papers are relevant because they address the issue of scale, analyzing malware signature generation and cloned application detection, respectively. However, we are interested in the dynamic analysis of Android applications, which gives insight into runtime behavior.

The work that comes closest to our work in scalable dynamic Android analysis is Crowdroid [6]. This paper creates a scalable analysis framework by crowdsourcing its data collection. It relies on users to run applications on their phones and send back information to a centralized server. While this is very similar to our companion application, Arkhunter, (described in the Discussion section), it only scales well when there are a large number of active users on the system. There may also be problems with users tampering with, or skewing the data based on their usage of the application. Andlantis avoids these issues by running the dynamic analysis on a collection of virtual Android devices instead of using a crowdsourced platform.

Similarly, Mahmood et. al developed a scalable dynamic analysis framework for analyzing Android applications in the cloud [17]. Their platform utilizes the robotium test automation framework to drive the user interfaces of the applications. Robotium allows for blackbox testing of Android applications through the development of a companion (testing) app that is installed alongside the testee application. The robotium platform requires the testee app to be signed in debug mode. Because production applications are rarely signed in debug mode, they must be resigned in order to work with robotium. Application signatures can be checked by the application during runtime, and applications may break or significantly reduce functionality if they detect they have been resigned. Because we are interested in the dynamic analysis of Android malware, we wish to avoid conflicting behaviors that occur when an application is modified. Andlantis can reap the benefit of operating on unaltered applications.

[4] and [14] detail Android dynamic analysis frameworks emphasizing coverage and target analysis. [4] achieved an activity coverage rate above 60% – over twice the coverage of human interaction with the same applications. [14] details a programmable framework for Android UI interaction that allows for flexibility in state exploration and easy access to program state.

Other works ([15], [5]) utilize Android's Monkey tool [3] to drive the UI of applications on an Android emulator. Monkey runs on the Android emulator, and exercises the application through user interface manipulation. Monkey does not require applications to be signed in debug mode, which is very important to our application of the Andlantis framework. Despite the fact that these dynamic analysis tools do not require application modification, their frameworks are not designed to scale to hundreds or thousands of nodes like Andlantis.

Although static analysis frameworks [13], [9] are scalable and provide valuable insight to Android application behavior, our work is focused on building a scalable dynamic analysis framework for evaluation of Android application behaviors during runtime.

## III. Architecture

As seen in Figure 1, our system is divided up into three main components: a malware analysis repository (FARM[20]), a distribution system built on top of a cluster management framework (Minimega[11]), and a behavioral analysis environment to capture forensic data. It runs on a large commodity cluster.

*1) Kane: Commodity Cluster for Malware Analysis:* In order to provide a safe, scalable, and reliable environment for the analysis of malware, we constructed Kane, a 520-node system consisting of reasonably priced commodity PCs. Each node is capable of hosting between 8 and 20 virtual machines, depending on the CPU and RAM requirements of the application and guest OS. The systems are intentionally diskless and PXE boot from a collection of experiment images stored on the cluster head node. Because there are no disks, even malware employing VM breakout techniques will have difficulty infecting the host machine permanently. A simple reboot will wipe the node clean, and any damage the malware could cause to other systems would be limited because our system's internal network has no connection out to the Internet.

*2) FARM: Malware collection and organization:* The Forensics Analysis Repository for Malware (FARM) is the primary input to our system. FARM stores known-bad, known-good, and unknown applications, as well as the results from automated analysis tools. Researchers and analysts can request a file to be analyzed by various systems, including Andlantis. In addition to managing the results from automated tools, FARM is able to leverage its large repository of malware to correlate information across samples. As mobile malware techniques approach the sophistication of PC-based malware, it will only become more important to be able to compare samples across different platforms for similar behavior and functionality.

*3) Andlantis Agent: Job control:* When an analyst selects a sample for examination, it is sent to the control node of our cluster. Here, a scheduling program locates a node with enough available resources to run the job. When a node is found, the Android APK is sent to the node, along with the instructions for dynamic analysis and emulator configuration. If no nodes exist with the resources necessary for the experiment, the scheduler will wait until a node becomes available.

*4) Minimega: large-scale VM management:* In order to control all the various parts of Kane, we use minimega, an open-source framework for deploying and managing networks of virtual machines across a large number of systems. Minimega was originally developed to study large-scale emulytics by creating networks of hundreds of thousands of networked VMs. It uses a mesh networking strategy to efficiently propagate commands and files to all or a subset of the nodes in our cluster, and provides a simple interface to control both the deployment of virtualized experiments, as well as the network topology crucial to isolating each experiment. Minimega supports Intel based Android VMs, which this work leverages to support thousands of concurrent analyses.

*5) Application interaction (Dynamic Analysis):* Initially, we developed Andlantis to launch highly parallelized virtual environments using the x86 Android architecture. However, we quickly discovered that the x86 images do not come with the Google Maps API, which many applications require in order to run. In order to retain this functionality, we added the Google Maps API to the x86 images. We also disabled the callbacks to the Google Bouncer service, which is intended to prevent malware from being installed on Android devices. Because Andlantis was developed to analyze malware, Google Bouncer checks often impeded our research, or would cause installs to hang if we did not have network connectivity.

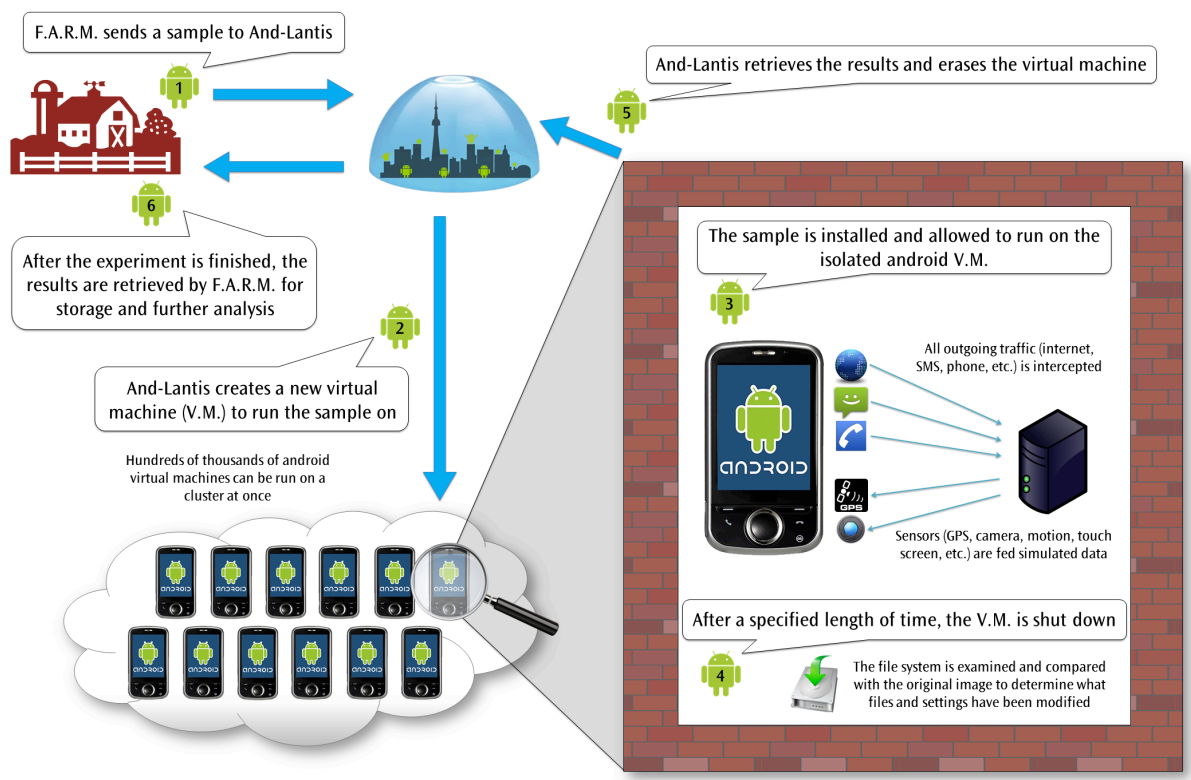

Fig. 1. "Andlantis architecture diagram"

Each emulator's GUI is controlled via an application utilizing the MonkeyRunner and AndroidViewClient frameworks to achieve easy and customizable UI exploration. By utilizing MonkeyRunner, we can control the behavior of the application and interact with the application's UI from outside of the Android code. The system will, depending on the kind of analysis requested, interact with an application's clickable, scrollable, and typable objects in various different patterns. Interaction patterns are recorded, along with application crashes, allowing us to determine what series of input events triggered a particular application behavior. In our current implementation, the UI stimulation attempts to visit as many features of the application as possible. If we reach a state in which there are no more UI elements to visit, we will visit the previous UI state, and try to traverse a new path by interacting with UI elements that have not yet been stimulated. Although this method may not cover all functionality of the application (e.g. it has trouble with text fields requiring specific information such as phone numbers and multitouch interaction), it allows us to observe many of the features of the application during runtime.

Each virtual experiment involves two virtual instances. One is the experimental instance on which the sample is installed and run. The other runs iNetSim[16], "a software suite for simulating common internet services in a lab environment". An isolated virtual network segment connects the two nodes. This iNetSim node will respond to network requests made by the emulator, spurring malicious samples to expose additional functionality. The host machine captures all network traffic on this network for later analysis.

### A. The Life of an Andlantis Job

When an analyst using FARM wants to analyze an application, it is retrieved from the repository and sent, along with various job-related metadata, to the Andlantis agent on the head node of the Kane cluster. Upon receiving the request, the Andlantis agent leverages minimega's scheduler to locate a node in the cluster with available CPU and RAM. This node, Node X, creates a new Android emulator and iNetSim instance. Once the emulator and iNetSim instances are booted, the head node sends the APK to be analyzed to Node X for deployment. Node X installs the APK into the emulator, launches the main activity of the contained application, and begins the automated interaction. After a period of time (currently configured for 5 minutes), the emulator is stopped, and its disk images are retained for analysis. During the forensics phase, the disk images are loop-mounted on the host node's filesystem and a simple filesystem comparison is run against the original Android image to determine what changes the target program made. The differences are categorized into created, modified, and deleted files. The created and modified files are copied from the disk image for future analysis. The files, metadata, and network traffic generated from the application's execution are then archived and by the head node, where they are made available to FARM for inspection.

## IV. EVALUATION

In order to evaluate the Andlantis system, we chose to run the set of applications from the Android Malware Genome Project[21]. The dataset contains 1261 malware samples collected between August 2010 and September 2011. This allowed us to both test our system at scale, and to evaluate the behavior of known Android malware in the controlled Andlantis environment.

### A. Job Scheduling

In order to send Android applications (.apk files) to nodes on our cluster for analysis, we wrote a scheduling program to manage the distribution. Using the minimega framework, the scheduler sends a broadcast message to all the nodes, requesting usage statistics (available RAM and CPU load average). All nodes with enough available resources are added to a queue. Each job is then sent to the next node in the queue. Once the queue empties, we repeat the processes. If no nodes meet the resource requirement, the scheduler waits for a configurable timeout period before retrying. As jobs are completed on the nodes, resources will be made available for new jobs to use. After a job has run for a specified time (generally 5 minutes), the scheduler stops the job, collects the resulting data from the appropriate node and cleans up the files used for the job. The entire scheduling routine is performed in parallel in order to maximize the efficiency of the process.

### B. Performance

We ran the Malware Genome Project set on a commodity cluster of 200 nodes. Each diskless node contains 12GB of RAM and a quad-core, 2.8GHz Intel Core i7 CPU. Because we were running at such a large scale, occasional problems with the nodes are to be expected. Hardware issues on 12 of the nodes rendered them useless for our analysis, so the actual test ran on 188 of the 200 nodes. During our experiment, 38 jobs did not complete because of errors related to the Android virtual machines. The leading cause of virtual machine problems was emulator time-out while waiting to acquire an IP address. This problem is nondeterministic and can be resolved by resubmitting the job until it successfully acquires an IP. In this experiment, we chose to simply abandon the job if it failed, although failure recovery logic can be easily added to the scheduler. Since the VM failure rate is only 3%, we believe it has only marginal impact on the efficacy of Andlantis and leave the addition of failure recovery logic to future work.

The entire run of 1261 samples was completed in 24 minutes, 25 seconds. Each Android application was stimulated for 5 minutes before the app was closed and results were collected. Theoretically, with 1261 applications each running for 5 minutes, and 188 nodes, we could run the entire experiment in just over 11 minutes, assuming three virtual machines per node. In practice, we found that running more than three virtual machines concurrently would saturate the available resources on our nodes. Obviously, this number is largely hardware dependent. The seemingly large discrepancy between the theoretical runtime and the actual runtime is largely due to wait times we have introduced in order to avoid timing issues. For example, we must wait for a period of time after the emulator boots before we check if it has acquired an IP address. We must choose between waiting too long and sacrificing performance, and not waiting long enough which results in failed jobs. Other wait times are introduced after kicking off the forensics analysis scripts. All of these wait times are configurable, and for our current implementation, we have a total of 220 seconds of wait time, excluding the 5 minutes we must wait for the stimulation of the Android application. In total, we must wait at least 520 seconds for each job to complete. Our theoretical runtime with no additional overhead now jumps to 19 minutes 23 seconds. This means that our system incurs approximately five minutes of overhead (roughly 0.25 seconds per job) associated with scheduling and transferring files to/from the nodes.

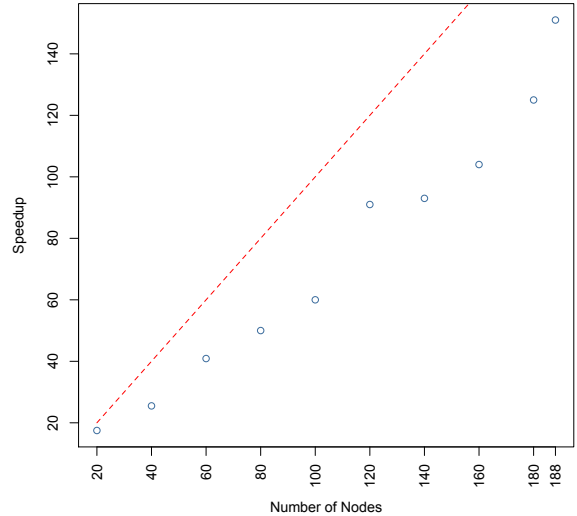

Fig. 2. "Speedup"

### C. Scalability

The Andlantis system naturally scales extremely well. As shown in Figure 2, we tested the Malware Genome Project set on Andlantis with a variable number of nodes (ranging from 20 to 188). The dotted line represents the theoretical speedup, while the dots represent our observed speedup with a given number of nodes. We can see that with the number of nodes tested, we have approximately linear strong scaling speedup. As the number of nodes increases, the cost of the dynamic analysis computation becomes less of an issue and the performance is bound by the communication costs to send and receive data to the nodes. This is shown in Figure 4. With 20 nodes, communication cost accounts for less than 4% of the total execution time. When we scale to 188 nodes, almost 25% of the execution time is used for communication with the nodes. The total runtime for the experiments is displayed in Figure 3. With 20 nodes, our total runtime was well over three hours and total runtime decreased as more nodes were added.

The use of Minimega facilitates automatically scaling to available resources by simply allowing new Minimega instances (nodes) to connect and disconnect. Minimega automatically includes participating nodes as available resources for scheduling. If new nodes are more powerful than the nodes currently in the network, the Andlantis scheduler will detect the available resources and be able to schedule more jobs on the new node than on existing nodes. Likewise, if the new node is less powerful, it may only be able to schedule one job (or none if it is a very low power computer). This system makes Andlantis very dynamic, as more nodes can be simply plugged into the network. Likewise, Andlantis is robust to issues such as node failures. If a node goes down, it will fail to respond to queries from the scheduler and therefore will never be scheduled. When the node comes back, it will automatically be detected by the system.

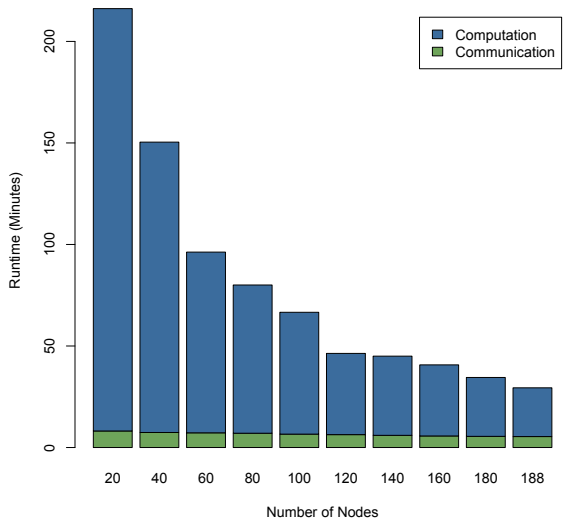

Fig. 3. "Performance of Andlantis"

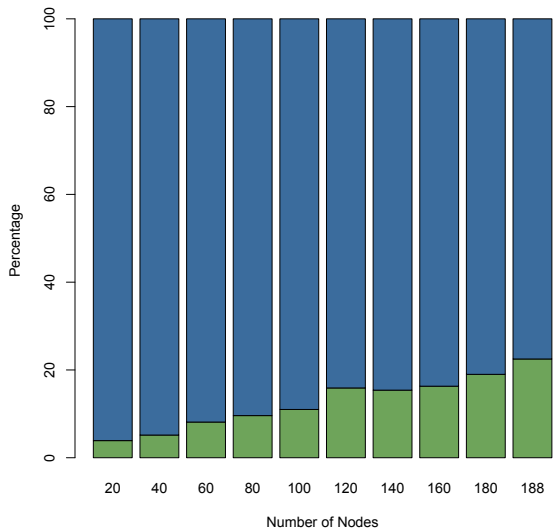

Fig. 4. "Percentage Breakdown"

## V. FINDINGS

The Andlantis system can be used to evaluate thousands of Android applications in a short amount of time. For our initial malware analysis run, we focused our efforts on the forensic evaluation of Android applications found in the Malware Genome Project set. Manual inspection was performed on applications found to have suspicious changes to their file system after five minutes of stimulation. Our analysis is not focused on finding new class of malware, rather observing the behavior of known Android malware. We present an overview a few families of malware detected by the Andlantis system, and detail our findings:

### A. AndroidOS/DroidKrungFu.A

The DroidKrungFu.A family of malware contains a trojan that sends personal information from the device to a command and control server. This information includes the contents of the memory card, IMEI, the mobile device number and the Android SDK version. In the applications in our analysis dataset, the DroidKrungFu.A malware was found packaged in an application advertised as a text reader. When the application was installed onto our device, it would also create a file named */data/media/0/txtbooks/legacy*. This file appears to be a normal data file, and is even placed in a directory titled *txtbooks* so as not to raise suspicion. However, the *legacy* file is actually and Android APK file, and we observed the installation of this APK on the device through our stimulation via dynamic analysis.

### B. AndroidOS/Anserver.A

Anserver.A is a class of Android malware targeted at stealing personal information from mobile devices. The malware will periodically phone home to various public blogging websites, where it finds the URLs of new command and control servers. Like the DroidKrungFu.A family, the Anserver.A malware is commonly found packaged in other applications. The most common instances of the malware contain a file named *anserva.db* which is an Android APK file. The encompassing application will then entice the user into installing the malware, promising added functionality for their device. In addition to allowing us to track the behavior of the Anserver.A malware, Andlantis system allowed us to observe the common methods the malware used to persuade the user into installing the malware.

### C. j.SMSHider

The j.SMSHider Android malware, like the other malware families discussed, is designed to gather personal information from Android devices. This information includes the phone number and GPS location of the victim. The malware will upload this information to a server roughly every five minutes. The application will also delete SMS messages to and from certain numbers. This malware differs from the previously examined families in that the malicious application gains root access on the device. This means that the application has privileged control of the device, usually reserved for the Android operating system, and not for installed applications. Our analysis allowed us to determine that the installed application had gained root access on the device, which immediately raised red flags. This gives us a large improvement over static analysis of APK files, as we are able to observe the state of the Android system during runtime.

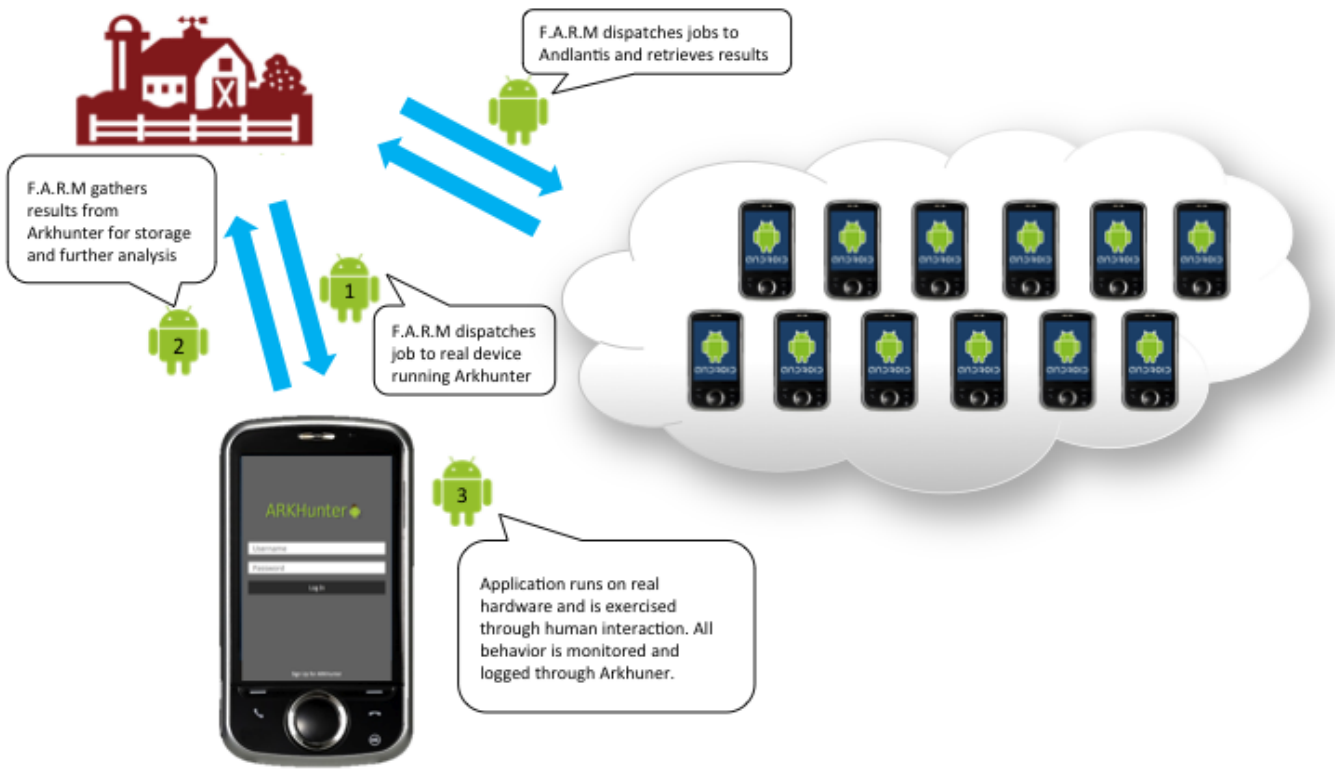

Fig. 5. "Arkhunter Interaction Diagram"

## VI. Discussion

Andlatis is able to perform large-scale dynamic analysis on Android applications in an emulated environment. It is the most scalable solution to date and is flexible enough to handle more advanced dynamic analysis techniques. Currently, we are able to process 1261 Android applications in about 25 minutes, which equates to about 3026 .apk files per hour (running them each for five minutes). This means that we can run the entire set of 1000000 applications available on the official Android marketplace, Google Play, in just under two weeks. Android system developers as well as application developers could use this tool to detect application crashes and trace the precise steps that resulted in the crash. This data could be used to improve the Android API, or to fix bugs existing in applications.

Although our platform is able to successfully stimulate applications and, in our case, gather forensics data, it is often useful to manually navigate through an application in order to increase coverage or activate specialized functionality within the application. To facilitate these capabilities, we developed a companion application to Andlantis for forensics analysis, which we named Arkhunter.

### A. Arkhunter

Arkhunter is a lightweight, standalone application that can be installed on any Android device. To use Arkhunter, the user simply opens the application and selects the "scan device" option. Once selected, Arkhunter will gather information on every file on the device. This information includes cryptographic hashes (MD5, SHA-1, SHA-256), the absolute file path of the files, and the status information of the files. The status information is equivalent to the information found in the *stat* system call.

Unfortunately, this information is not easily obtainable from standard Java libraries. To gather this information, we used the Android Native Development Kit (NDK) to write a C function to make the stat call and return the relevant information back to the Java application. This consists of the ID of the device containing the file, inode number, permissions, number of hard links, user ID of the owner, group ID of the owner, device ID if the file is a special file, filesize, blocksize, number of blocks allocated, time of last access, time of last modification and time of last status change. This information is then packed into a JSON object that can be saved locally or exported to a remote server. The format of this data is identical to the format of the Andlantis forensics data, which enables researchers to easily integrate data from both Andlantis and Arkhunter. In our testing, this data was analyzed by the FARM system. As displayed in Figure 5, Arkhunter is easily added to FARM along side the Andlantis system in order to perform more customized forensics analysis of Android applications on a real device.

## VII. LIMITATIONS AND FUTURE WORK

While we have achieved a significant step forward in scalable Android application analysis, this work does have its limitations. It is implied by the nature of dynamic analysis, but bears repeating, that the quality of results gained by our system is dependent on the actual percentage of an app's code that is executed, and therefore the means by which this execution occurs.

In our system, this is the UI exploration mechanism, which currently uses a rather naive algorithm of interacting with random clickable and typable UI elements. With more intelligent algorithms, an increase in coverage can be attained. However, there are likely significant diminishing returns from this, since many Android apps have behaviors that simply cannot reasonably be activated without human intervention. The behaviors include application-specific login prompts, applications that dynamically load content or code from network resources, etc.

In the future, our first step will be to tackle the above problem most directly by developing a means by which to measure code coverage during execution in Android applications. One could hypothesize various methods for doing this, such as application rewriting, or through the use of the Java debugging framework, but the pros and cons of these methods need to be evaluated. With this accomplished, the coverage data can then be used to evaluate various improvements to our exploration and interaction mechanisms. One of the most important of these improvements is the use of Google's "Goldfish" Android emulator, which will allow us to manipulate simulated network parameters, send and receive text messages, spoof hardware sensor data, and so on.

Additionally, we recognize that due to the fragmented nature of Android, applications will behave differently based on the API level and architecture of the device they are running on. While our emulators run Android 4.2 on the x86 architecture, we could get more complete data by supporting other variants, which are the most common targets for malware.

## VIII. CONCLUSION

We have developed Andlantis, a highly scalable system for the dynamic analysis of Android applications. This system, at its current state, has the capacity to process over 3000 applications per hour, and can greatly exceed this number with additional hardware. Andlantis is robust to node failures, and has a very low job failure rate of just 3%.

We have shown the ability to evaluate the forensic footprint left by three common families of mobile malware (Droid-KrungFu.A, Anserver.A and j.SMSHider) and are able to further evaluate runtime behaviors and network traffic of these malicious applications. The ability to evaluate thousands of applications in parallel allows forensics experts to focus their time on the more dangerous, or more forensically interesting applications.

## ACKNOWLEDGEMENTS

We would like to express a special thank you to Michael Kunz and Natalie Roe for their work on the Arkhunter project and to Steven Barker, Miles Carabill and Joan Hong for their contributions to this project during their Center for Cyber Defenders (CCD) summer internships.